\documentstyle[12pt]{l-aa}
\input psfig.sty

\def\msun{$M_{\odot}$}
\def\db{$\Delta{B}$}
\begin{document}
   \thesaurus{08         
              (08.04.1;  
               08.06.3;  
               08.15.1;  
               08.22.1)} 
\title{An investigation of the revised CORS method based on theoretical models}

   \author{V. Ripepi
           \inst{1,2}
   \and    G. Russo
          \inst{3,4}
   \and    G. Bono
          \inst{5}
   \and    M. Marconi
	  \inst{1,2}	 
          }
          
   \offprints{V. Ripepi}

\institute{Dipartimento di Fisica, Universit\`a di Pisa, Piazza Torricelli 2, 56100 Pisa, Italy
\and Osservatorio Astronomico di Capodimonte, Via Moiariello 16, 80131 Napoli, Italy
\and Dipartimento di Fisica, Universit\`a della Calabria, Arcavacata di Rende, 87036 Cosenza, Italy	
\and CDS Universit\`a di Napoli Federico II, Complesso Monte S. Angelo, 80126 Napoli, Italy
   \and Osservatorio Astronomico di Roma, Via Frascati 33, 00040 Monte Porzio Catone, Italy 
}


   \maketitle

\begin{abstract}

Direct measurements of Cepheid radii are a key for understanding the physical  
structure of these variables, and in turn for constraining their pulsation 
properties. In this paper we discuss the numerical experiments we performed
for testing the accuracy of Cepheid radii obtained by adopting 
both the pure Baade-Wesselink (BW) method and the revised CORS method, 
as well as the consistency of the physical assumptions on which these 
methods are based. We applied both the BW and the revised CORS methods 
to the synthetic light, color and radial velocity curves predicted by 
Cepheid full amplitude, nonlinear, convective models at solar chemical 
composition. 

We found that these methods systematically either underestimate or 
overestimate "theoretical" radii if radius determinations are 
based on optical $(BVR)$ bands or on $(VJK)$ bands, respectively.  
At the same time, current simulations suggest that CORS radii are in 
very good agreement with "theoretical" radii if the surface brightness 
is calibrated by adopting a bidimensional fit of atmosphere models 
which accounts for temperature, gravity, and bolometric correction 
variations along the pulsation cycle. 

Finally, a slight discrepancy between "computed" and "theoretical" 
radii of a Bump Cepheid supports the exclusion of these pulsation phases 
in both BW and CORS analyses. In fact, we found that the assumption 
of quasi-static approximation is no longer valid during the pulsation 
phases in which appears the bump.

\end{abstract}

%
%

\section{Introduction}

The Baade-Wesselink (BW, Baade 1926; Wesselink 1946) method on the 
basis of luminosity, color, and radial velocity variations along the 
pulsation cycle provides the key opportunity to estimate both radii and 
distances of variable stars. Both the physical assumptions on which 
this method relies and the intrinsic drawbacks have been thoroughly 
discussed in the literature (Oke et al. 1962; Gautschy 1987; 
Bono et al. 1994; Butler et al. 1996).  

Different approaches have been suggested for improving both accuracy 
and consistency of the BW method: \par
a) the radius variations are estimated by adopting a maximum likelihood 
method (Balona 1977; Laney \& Stobie 1995) which accounts for observational 
errors.\par
b) The use of light and velocity variations together with two color 
indices to account for both temperature and gravity changes along the 
pulsation cycle (Caccin et al. 1981; Sollazzo et al. 1981; 
Onnenbo et al. 1985, the CORS group). However, the CORS method needs 
photometric calibration and a good sampling of both light and velocity 
curves for evaluating temperatures and gravities. \par
c) The use of a surface brightness relation (Barnes \& Evans 1976; 
Gieren et al. 1989; Fouqu\'e \& Gieren 1997; Gieren et al.  
1997, hereinafter GFG). However, this method needs accurate calibration 
of the  surface brightness parameter in order to provide simultaneous 
estimates of radii and distances. \par
d) A detailed comparison between theory and observations brought out 
that Cepheid radii and distance determinations should be based 
on atmosphere models constructed by adopting a microturbulence 
velocity of the order of 4 km/sec (Bersier et al. 1997). 
Thus confirming the result originally pointed out by 
Lub \& Pel (1977) and Pel (1978).\par 
e) In a recent paper Ripepi et al. (1997, hereinafter RBMR) revised the 
CORS method by including the surface brightness calibration suggested 
by Barnes \& Evans (1976). This method has been applied to a large 
sample of Galactic Cepheids and the radius estimates they obtained 
are in fair agreement with previous evaluations.\par
f) Krockenberger et al. (1997) adopted a Fourier analysis of both 
light and velocity curves to account for individual measurement 
errors.\par

On the basis of these developments and of accurate photometric 
and spectroscopic data it has been suggested that the most recent 
estimates of Cepheid radii are affected by very small internal 
errors (Di Benedetto 1997; GFG). 
Moreover, in a recent investigation based on new Cepheid models 
Bono et al. (1998, hereinafter BCM) settled a long-standing 
discrepancy between theoretical and empirical Period-Radius (PR) 
relations (Laney \& Stobie 1995)\footnote{
The theoretical PR relations provided by BCM were estimated by 
adopting a large set of full amplitude, nonlinear, convective models
which cover a wide range of stellar masses ($5 \le$M/\msun$\le 11$), 
and effective temperatures ($4000 \le T_e \le 7000$ K).
The models were constructed by adopting three chemical compositions 
which are representative of Galactic ($Y=0.28$, $Z=0.02$), Large 
Magellanic Cloud (LMC, $Y=0.25$, $Z=0.008$), and Small Magellanic 
Cloud (SMC, $Y=0.25$, $Z=0.004$) Cepheids. On the basis of predicted 
periods and radii, BCM derived analytical PR relations at fixed 
chemical composition of the type $logP=\alpha+\beta logR$.}. 
In fact, they found very good agreement between theory and observations 
in the period range $0.9 \leq \log P \leq 1.8$. However, they also 
found that outside this range, at both shorter and longer periods 
theoretical predictions attain intermediate values between empirical 
radii estimated by adopting different BW methods and/or photometric 
bandpasses. 

Even though, it has been recently suggested that period and radii of 
Cepheids obey to a {\em universal} PR relation, theoretical predictions 
support the evidence that both the slope and the zero point of this 
relation depend on metallicity (BCM). Moreover, it has also been 
estimated that the accuracy of Cepheid radii based on infrared colors 
is of the order of 3\% (GFG) and therefore the metallicity dependence, 
if any, should have already been detected. However, preliminary results 
(Laney 1999a,b) based on a large sample of Galactic and MC Cepheids for 
which multiband photometric data are available seem to support theoretical
predictions, and indeed he found that the radii of MC Cepheids are, at one
$\sigma$ level, systematically larger than the radii of the Galactic ones.  

A similar discrepancy has been found between theoretical and empirical 
estimates of the Cepheid intrinsic luminosity.   
In fact, recent theoretical investigations support the evidence that 
the Cepheid PL relation depends on the metallicity, since at fixed 
period metal-rich Cepheids are fainter than metal-poor ones 
(Bono et al. 1999a, hereinafter BMS; Bono et al. 1999b, hereinafter BCCM). 
However, these predictions are at odds with current empirical estimates 
based on the BW method or on other approaches, since they show that the PL 
relation is either unaffected by the metallicity (GFG), or it presents 
a mild dependence but with an opposite sign, i.e. metal-rich Cepheids 
seem to be brighter than metal-poor ones (Sasselov et al. 1997; 
Kennicutt et al. 1998). 

The main aim of this investigation is to test both physical and 
numerical assumptions adopted for developing the revised CORS method 
by performing a set of numerical experiments based on theoretical 
light, color and radial velocity curves.   
The pulsation models and the static atmosphere models adopted for 
transforming theoretical observables into the observative plane are 
discussed in \S 2. 
In section 3 we briefly summarize the leading equations on which
the revised CORS method is based and then we describe the approach 
adopted for testing the method. The results of the numerical 
experiments we performed are presented in sections 4.1 and 4.2, 
together with a detailed analysis of the dependence of radius 
estimates on the photometric bands currently adopted. 

A new calibration of the surface brightness based on atmosphere 
models, which accounts for both temperature and gravity changes 
of classical Cepheids, is discussed in \S 4.3. In this section 
the improvements in CORS radii obtained by adopting the theoretical 
instead of the empirical calibration are also presented together 
with the limits of the quasi-static approximation close to the bump 
phases. A brief discussion on future developments closes the paper.

\section{The synthetic curves}

In order to test both the accuracy and the consistency of the revised 
CORS method we adopted the observables predicted by hydrodynamical models
of variable stars. The reader interested in a detailed discussion on the 
physical assumptions adopted to construct these models and on the 
comparison between theory and observations is referred to BCM, BMS, 
and BCCM. Among the different sequences of nonlinear models we selected 
canonical models\footnote{The canonical models were constructed by 
adopting a mass-luminosity relation based on evolutionary tracks 
which neglect the convective core overshooting during hydrogen 
burning phases (Castellani et al. 1992).} at solar chemical composition 
(Y=0.28, Z=0.02) and stellar masses ranging from 5 to 11 $M_\odot$. 
At fixed stellar mass we generally selected three models which are 
located in the middle of the instability strip as well as close to 
the blue and the red edge. The period of the selected models roughly 
ranges from 3.5 to 106 days.  
The input parameters and the pulsation periods are summarized in Table 1
which gives, from left to right, (1) the model identification, (2) the 
stellar mass, (3) the luminosity, (4) the effective temperature, 
(5) the nonlinear time average radius along a full pulsation cycle, 
(6) the nonlinear pulsation period. 

\begin{table}[h]
\caption{Physical properties of the selected Cepheid models \label{tabella1}}
\begin{tabular}{rrccrr}
\hline
\hline\noalign{\smallskip}
model &	Mass            & Luminosity &	  $T_e$ & Radius & Period \\
      & $\rm M/M_{\odot}$ & $\rm \log{L/L_{\odot}}$ & K & $\rm R_{\odot}$ & Days \\
\hline\noalign{\smallskip}
5m1  &  5    & 3.07  &  5800  &   34.3  &  3.5231     \\
5m2  & 	5    & 3.07  &  5600  &   36.7  &  3.9569     \\
7m1  & 	7    & 3.65  &  5300  &   81.0  &  12.1307    \\
7m2  & 	7    & 3.65  &  5000  &   91.3  &  14.7877    \\
7m3  & 	7    & 3.65  &  4800  &   97.3  &  16.8658    \\
9m1  & 	9    & 4.00  &  4900  &   141.1 &  27.2763    \\
9m2  & 	9    & 4.00  &  4700  &   152.8 &  31.3729    \\
9m3  & 	9    & 4.00  &  4500  &   164.5 &  36.0966    \\
11m1 & 	11   & 4.40  &  4800  &   230.2 &  59.7157    \\
11m2 & 	11   & 4.40  &  4300  &   282.7 &  86.3676    \\
11m3 & 	11   & 4.40  &  4000  &   310.6 &  106.5060   \\
\noalign{\smallskip}
\hline
\hline
\end{tabular}
\end{table}

Theoretical observables have been transformed into the observational 
plane by adopting the bolometric corrections (BC) and the 
color-temperature relations by Castelli et al. (1997a,b). 
We assumed $M_{Bol}(\odot)$=4.62 mag.  
The main difference between the static atmosphere models constructed by
the quoted authors and the grid of models computed by Kurucz (1992) is 
that overshooting was neglected. In fact, they found 
that for temperatures and gravities typical of the Cepheid instability 
strip both the color indices and the Balmer profiles of the models 
constructed by neglecting overshooting are in better agreement 
with observational data. 
Unfortunately the set of atmosphere models provided by Castelli et al. 
(1997a,b) was constructed by adopting a fixed value of microturbulence 
velocity $\xi =2 km s^{-1}$.  Even though it has been recently suggested 
by Bersier et al. (1997) that theoretical colors based on atmosphere models 
which adopt higher microturbulent velocities are in better agreement with 
observational data, we plan to investigate the dependence on this parameter 
as soon as homogeneous sets of atmosphere models constructed by adopting 
different $\xi$ values become available.  

\begin{figure*}
\psfig{figure=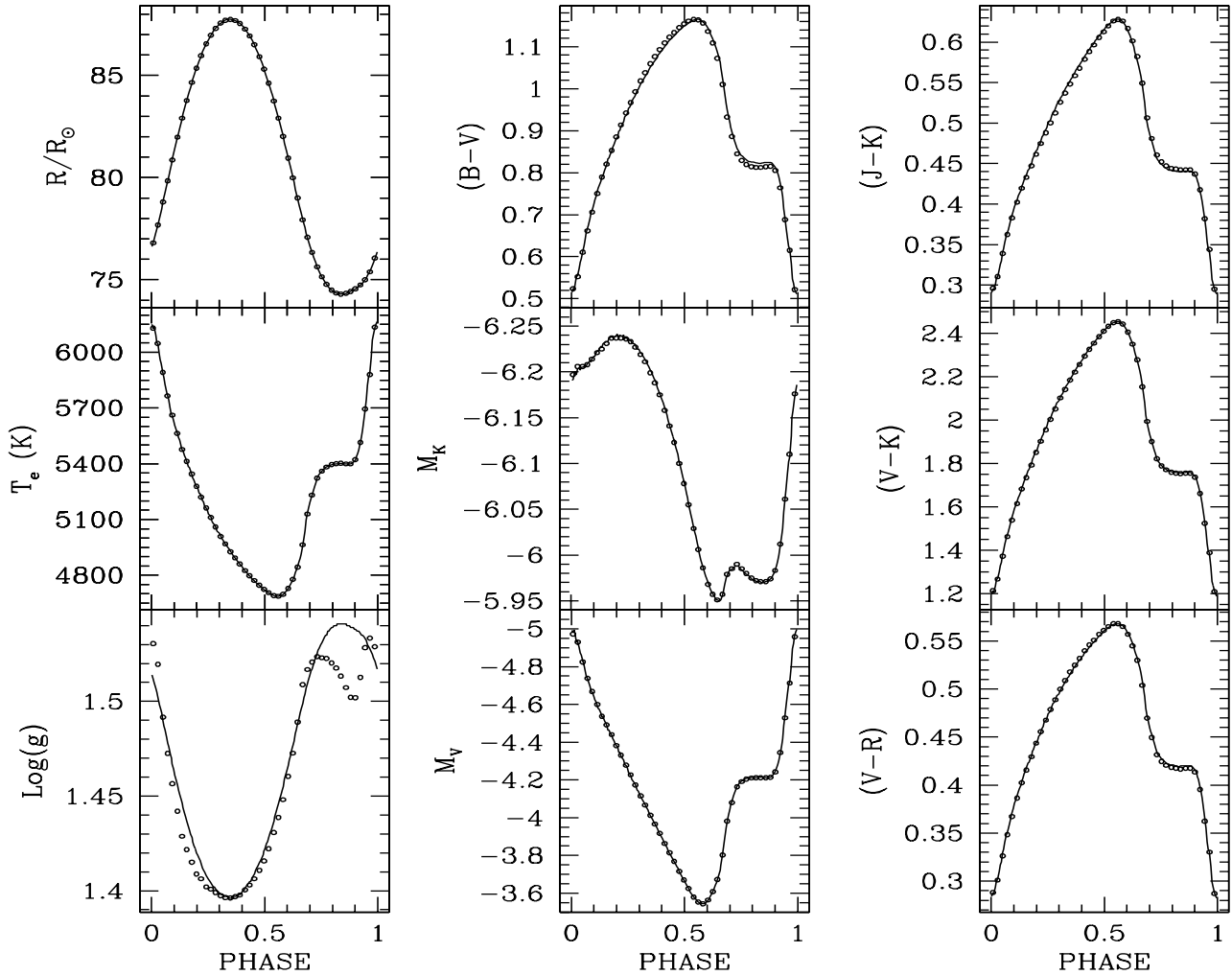,height=15cm,width=18cm}
\caption{Variations along a full pulsation cycle of several theoretical 
observables for the model at 7 \msun~ and $T_e=5300$ K. This model present 
a well-defined bump soon after the phase of luminosity maximum. 
Solid lines and dots display magnitudes and colors transformed by adopting 
static and effective gravities respectively.\label{fig1}}
\end{figure*}

To account for the effect of the gravity on both magnitudes and colors,   
the luminosity and temperature variations along the pulsation cycle
have been transformed by adopting static and effective gravities\footnote{
The static gravity is defined as $g_{stat}=GM/{R^2}$, while the 
effective gravity as $g_{eff}=g_{stat}+{du/dt}$, where $u$ is the 
radial velocity. For a detailed discussion concerning the dependence on 
the effective gravity the interested reader is referred to 
Lub \& Pel (1977) and to Bersier et al. (1997).}. 
In the following the models transformed by adopting $g_{stat}$ 
and $g_{eff}$ will be referred to as "static" and "effective" 
models, respectively. For each model we have taken into account 
two magnitudes -$V$, $K$- and four colors, namely 
$(B-V)$, $(V-R)$, $(V-K)$, and $(J-K)$.  
Figure 1 shows these curves together with the variations of radius, 
effective temperature, and gravity for the model at 7 \msun~ and 
$T_e=5300$ K. The curves plotted in this figure show quite clearly that 
both magnitude and colors present a negligible dependence on gravity. In 
fact, even though static and effective gravities attain different values 
along the cycle and present a difference of the order of 0.05 dex 
close to the bump phases, the two synthetic curves are almost identical
(for a detailed analysis of the dependence of bump Cepheids on static and 
effective gravities see \S 4).

\section{Test of the revised CORS method}

In the following we briefly summarize the main features of the CORS method. 
The reader interested in a comprehensive discussion on the adopted physical 
and numerical assumptions is referred to RBMR and references therein. 
The CORS method relies on the definition of surface brightness $S_V$:

\begin{equation}
{ S_V = V + 5 \cdot \log {\alpha} } 
\end{equation} 
\hspace*{1.2cm} $\Updownarrow$
\begin{equation}
M_V - S_V + 5 \cdot \log (R/R_{\odot}) = cost. \label{eqn2}  
\end{equation} 

\noindent 
The solution is found by differentiating Eq.~\ref{eqn2} with respect 
to the phase, then by multiplying the result for a color index, e.g. 
$(B-V)$,  and eventually by integrating along the full cycle. 

Since the radial velocity is tightly connected with the pulsation velocity
according to:

\begin{equation}
{  \dot{R}(\phi)=-p \cdot P \cdot u(\phi)} \label{vel} \,
\end{equation} 

\noindent
we obtain the following equation:

\begin{eqnarray}
\lefteqn{  a \int_0^1 \log \bigl\{ 
R_0(\phi)-p \cdot P \int_{\phi_0}^{\phi} u(\phi^{\prime}) \cdot 
 d {\phi^{\prime}}  
\bigr\}   
\dot{(B-V)}(\phi) \cdot d \phi +} \nonumber \\
& & -B+\Delta B = 0   \label{cors}
\end{eqnarray} 

\begin{equation}
{  B=\int_0^1 (B-V)(\phi) \cdot \dot{V}(\phi) \cdot d \phi} \,
\end{equation} 

\begin{equation}
{  \Delta B=\int_0^1 {(B-V)}(\phi) \cdot \dot{S_V}(\phi) \cdot d \phi} 
\label{eqdb} \,
\end{equation} 

\noindent 
where $\phi$ is the phase, $P$ is the pulsation period, $R_0$ is the 
radius (in solar units) at a given phase ${\phi}_0$, $u$ is the radial 
velocity, $p$ is the radial velocity projection factor 
(Parsons 1972; Gieren et al. 1989; Sabbey et al. 1995) 
and $a$ is a constant equal to $5 \cdot log_e 10$.   

The numerical solution of Eq.~\ref{cors} supplies the unknown quantity 
$R_0$. In order to evaluate the radius as a function of the phase we adopt 
Eq.~\ref{vel} and finally the mean radius is estimated by averaging along 
the radius curve. By neglecting the \db~term in Eq.~\ref{cors}, 
we obtain the pure Baade-Wesselink method which requires a radial velocity,
a magnitude and a color curve for each individual variable.
A more precise radius determination can be obtained by including the 
\db~term. In fact, Sollazzo et al. (1981) and RBMR demonstrated that 
the inclusion of this term improves the accuracy of radius estimates, 
provided that $S_V$ is evaluated at each pulsation phase. 
The \db~term was included in the original CORS method 
(Sollazzo et al. 1981) by adopting the empirical photometric 
calibration of the Walraven system provided by Pel (1978), 
and in the revised CORS method (see  \S 2.3 in RBMR) by adopting 
the empirical calibration of the {\em reduced} surface brightness 
$F_V$ ($S_V = const -10 \cdot F_V$) as a function of $(V-R)$, 
provided by Barnes \& Evans (1976). 
This change allowed RBMR to apply the CORS method to a large sample 
of Cepheids for which photometric data in the conventional $BVRI$ 
bands were available.  
It is worth underlining that both the original and the revised CORS 
method do require two color curves but the latter method, thanks to 
the new calibration, can supply radius estimates of Cepheids for 
which are available two different pairs of Johnson/Cousins color 
indices. 

In order to test  
the accuracy of the \db~term evaluation we apply the two previous 
approaches to synthetic light, color, and radial velocity curves.  
In particular, we adopted theoretical periods and the synthetic 
curves, covered with 125 points, were fitted with Fourier series 
which include up to 31 terms (15 sine, 15 cosine plus a constant term) 
and eventually the quantities B and $\Delta{B}$ were evaluated as well.  
We adopted a large number of both points and Fourier terms, since we are 
interested in testing the accuracy of the revised CORS method by adopting 
theoretical templates which should not be affected, within the intrinsic 
uncertainties, by systematic and/or deceptive errors.  
 
Since the modified CORS method requires an empirical estimation of the 
surface brightness as a function of a color, in this investigation we 
applied the calibrations provided by Fouqu\'e \& Gieren (1997) on the 
basis of stellar angular diameter measurements collected by 
Di Benedetto (1993) and Dyck et al. (1996), i.e.:

\begin{eqnarray}
 F_{V_0} &=& 3.947-0.380 {(V-R_J)}_0  \label{eqn1} \\
 F_{V_0} &=& 3.947-0.131 {(V-K)}_0  \\
 F_{K_0} &=& 3.947-0.110 {(J-K)}_0 
\end{eqnarray}

\noindent
where $F$ is the {\em reduced} surface brightness and $R_J$ is the Johnson
$R$ band. However, it is worth noting that the $R$ photometric bandpass 
adopted by Castelli et al. (1997a,b) is the Cousins band. Therefore to 
account for the color difference between $(V-R_J)$ and $(V-R_C)$ the slope 
in Eq.~\ref{eqn1} has to be replaced with 0.521 according to the 
transformation provided by Bessel (1979). 
The uncertainty on this color transformation, due to a twofold 
fortunate circumstance, has negligible effects on radius estimates. 
In fact, as discussed by RBMR, only the slope of the $F_V$ versus color 
calibration is taken into account in the revised CORS method, since 
the zero point does not affect the derivatives.  
On the other hand, a change of the order of 30 \% in the slope of 
Eq.~\ref{eqn1}, due to the additive nature of the \db~term, causes a change 
of only 4 \% in the radius estimates. 

A similar calibration -$F_V$ versus $(V-K)$- was originally suggested 
by Di Benedetto (1995). However, for applying the revised CORS method to 
different colors, we adopted the multiband calibrations provided by 
Fouqu\'e \& Gieren (1997). 
We emphasize once again that the modified CORS method adopts one magnitude 
and two color curves (cases 2, 3, 5 below), whereas the pure 
BW method adopts one magnitude and one color curve (cases 1, 4, 6 below). 
On the basis of the selected bands we investigated the following 
combinations of magnitudes and colors:  

\begin{enumerate}
\item   
$V$, $(B-V)$

\item	
$V$, $(B-V)$, $(V-R)$ 
	
\item	
$V$, $(B-V)$, $(V-K)$ 
	
\item	
$V$, $(V-K)$ 
	
\item	
$K$, $(V-K$), $(J-K)$ 

\item	
$K$, $(J-K)$ 

\end{enumerate}

\noindent 
These bands were selected because they are quite common in the current 
literature, and also because they give a proper coverage of both 
optical and NIR wavelenghts. 


\section{Results}

\subsection{Dependence of the \db~term on photometric bands}

The main aim of the present analysis is to provide tight constraints 
on the \db~term adopted in the CORS method. This term quantifies the 
area of the loop performed by the variable in the $S_V$-color plane 
and therefore provides an estimate of the failure of the BW assumption 
that phases of equal color are also phases of equal temperature.
In fact, the area of the loop described by the Cepheid in this plane 
supplies fundamental information on the variation of both effective 
gravity and effective temperature values along the pulsation cycle
(Caccin et al. 1981). 

\begin{figure}
\psfig{figure=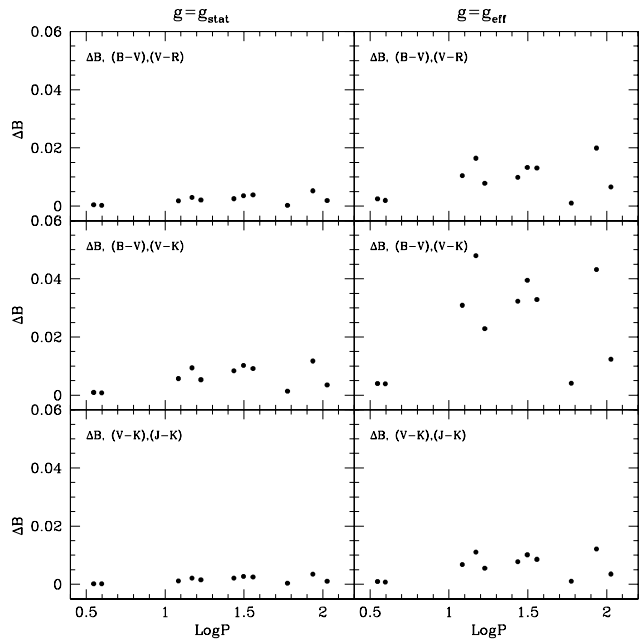,height=9cm}
\caption{\db~ values as a function of the logarithmic period for the 
whole sample of Cepheid models. The left panels show the models transformed 
into the observational plane by adopting the static gravity, while the right 
ones the models transformed by adopting the effective gravity. 
\label{fig2}}
\end{figure}

Fig.~\ref{fig2} shows the $\Delta{B}$ values we obtained by adopting  
the selected magnitude and color combinations (see labels) for both 
static (left panels) and effective (right panels) models. The first 
interesting outcome is that $\Delta{B}$ values  
attained by static models are systematically smaller than the values of
the effective models. This can be easily explained by the fact that the 
area of the color-color loops performed by effective models is larger 
than that of the static ones. This difference is caused by the sudden 
changes in the acceleration term (see \S 2) during the phases of rapid 
expansion and/or contraction. This larger excursion implies not only a 
difference in the area of the loop but also a change of its shape.   

The $\Delta {B}$ values of effective models present substantial differences  
between different color pairs, and indeed the values attained for  
$[(V-K), (J-K)]$ colors are at least a factor of three smaller than the 
values for $[(B-V), (V-K)]$ colors. 
However, we note that radius evaluations based on colors which present 
large $\Delta {B}$ values are not {\em a priori} more reliable than the 
evaluations based on colors with small $\Delta {B}$ values. In fact, 
in the following we show that radii based on $[(V-K), (J-K)]$ colors are 
more in agreement with theoretical radii than the radii based on 
$[(B-V), (V-K)]$ colors. At the same time, large $\Delta {B}$ values do not 
{\em a priori} imply that the CORS method is a major breakthrough in 
radius evaluations when compared to the BW method. In fact, radius 
estimates based on the BW method in $[V, (V-K)]$ and on the CORS method in 
$[V, (B-V), (V-K)]$ are in very good agreement with theoretical radii, and 
the discrepancy for both of them is smaller than 10\%. 

\begin{figure}
\psfig{figure=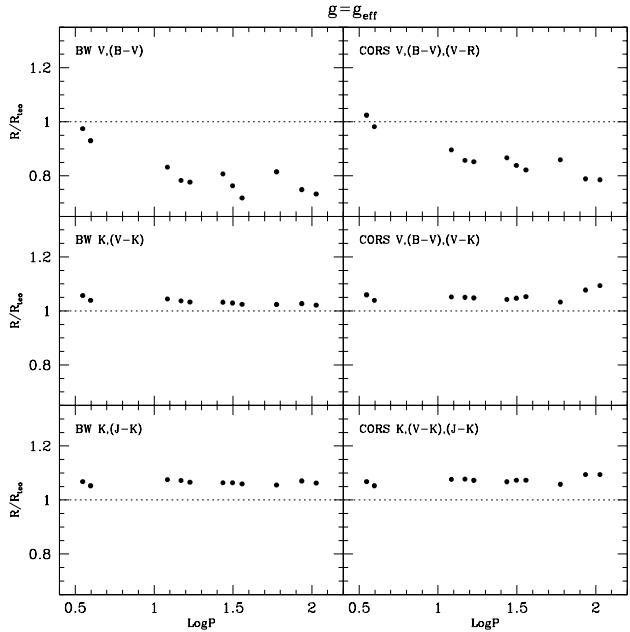,height=9cm}
\caption{Ratio between "computed" and "theoretical" radii as a function of 
the logarithmic period. The left panels display the radius evaluations based 
on a pure BW method, while the right ones the radius evaluations based on
the revised CORS method. The radius estimates plotted in this figure 
refer to models transformed by adopting $g=g_{eff}$. \label{fig3}}
\end{figure}

\subsection{Dependence of radius estimates on photometric bands}

The radius estimates of effective models in different photometric 
bands are plotted in Fig.~\ref{fig3}. Left and right panels show  
the radii evaluated by neglecting (pure BW method) and by including 
(revised CORS method) the $\Delta{B}$ term respectively. 
To evaluate the accuracy of radius estimates based on different 
methods, in this figure we plotted the ratio between "computed" 
and "theoretical" radii. Data plotted in the left panels show quite 
clearly that BW estimates based on $[V, (V-K)]$ and $[K, (J-K)]$ bands are 
in very good agreement with theoretical radii, and indeed the discrepancy 
is systematically smaller than 10\%. On the other hand, the radius 
evaluations in $[V, (B-V)]$ are systematically smaller than the 
predicted ones and the discrepancy is of the order of 30\% close to 
$\log P \approx 1.6$. Thus confirming the empirical evidence originally 
pointed out by Welch (1994) and by Laney \& Stobie (1995) that the use 
of $(V-K)$ colors or infrared bands ensures more accurate measurements 
of Cepheid radii.  

This result is further strengthened by radius estimates obtained 
by means of the revised CORS method (left panels). In fact, the 
discrepancy between "computed" and "theoretical" radii is generally 
smaller than 10\% when both $[V, (B-V), (V-K)]$ and $[K, (V-K), (J-K)]$ 
bands are adopted. At the same time, it is worth noting that radius 
determinations based on optical bands -$[V, (B-V), (V-R)]$- present 
a discrepancy smaller or equal to 20\% over the entire period range. 
The results of our numerical experiments suggest that by adopting 
NIR bands the radii evaluated through the revised CORS method 
present on average the same accuracy of the radii based on the 
pure BW method. However, the former method supplies more accurate 
radius determinations than the latter one when optical bands are 
adopted. Thus supporting the plausibility of physical and numerical 
assumptions adopted in the revised CORS method.

\begin{figure}
\psfig{figure=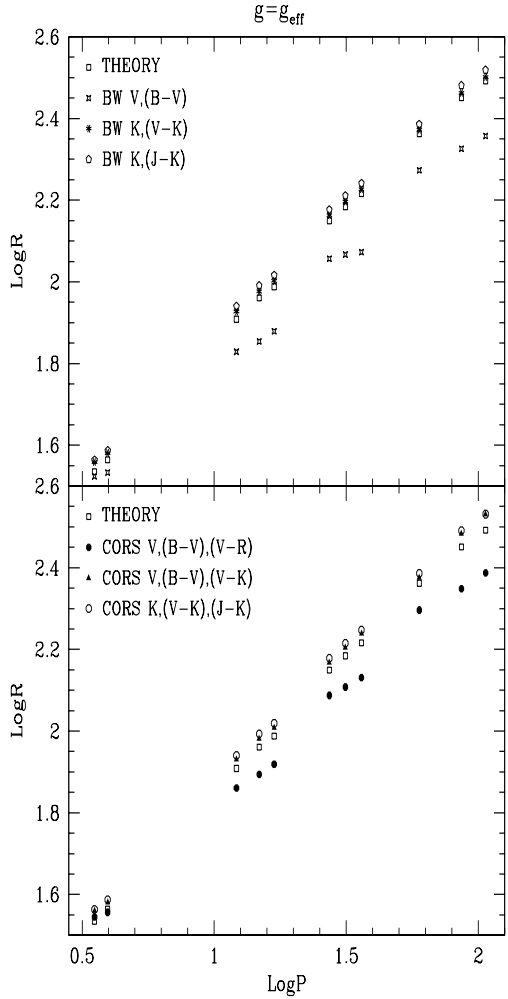,height=14cm,width=10cm}
\caption{"Computed" and "theoretical" Cepheid radii as a function of the 
pulsation period in a $log-log$ plane. The top panel shows the radius 
estimates obtained by adopting a pure BW method, while the radii 
plotted in the bottom panel by adopting the revised CORS method. 
Radius estimates based on different magnitudes and/or colors are 
displayed with different symbols. \label{fig4}}
\end{figure}

We will now focus our attention on the choice of the 
photometric bands which should be adopted for providing 
accurate radius determinations. 
Fig.~\ref{fig4} shows the comparison between "computed" and "theoretical" 
radii in the $log P-log R$ plane. Data plotted in the top and in the 
bottom panels display radius estimates based on the pure BW and on the 
revised CORS method respectively. The main outcomes of this comparison 
are the following:
1) the slope of the PR relation, as already noted by Laney \& Stobie (1995),
becomes steeper when moving from optical to NIR bands. 
2) Radius estimates based on NIR/optical bands overestimate/underestimate 
theoretical radii.  
These results apply to radius evaluations based on the pure BW method 
and on the revised CORS method, thus supporting the evidence that 
this "photometric drift" is not an artifact of the method adopted for 
estimating the radius. At the same time, data in the bottom panel of
Fig.~\ref{fig4} suggest that radii obtained by averaging the estimates 
in the $[V, (B-V), (V-R)]$ and in the $[K, (V-K), (J-K)]$ bands are 
much less affected by systematic errors than the radius evaluations  
only based on NIR bands or on optical bands.  

\subsection{A theoretical estimate of the $\Delta B$ term}

\begin{figure}
\vbox{
\psfig{figure=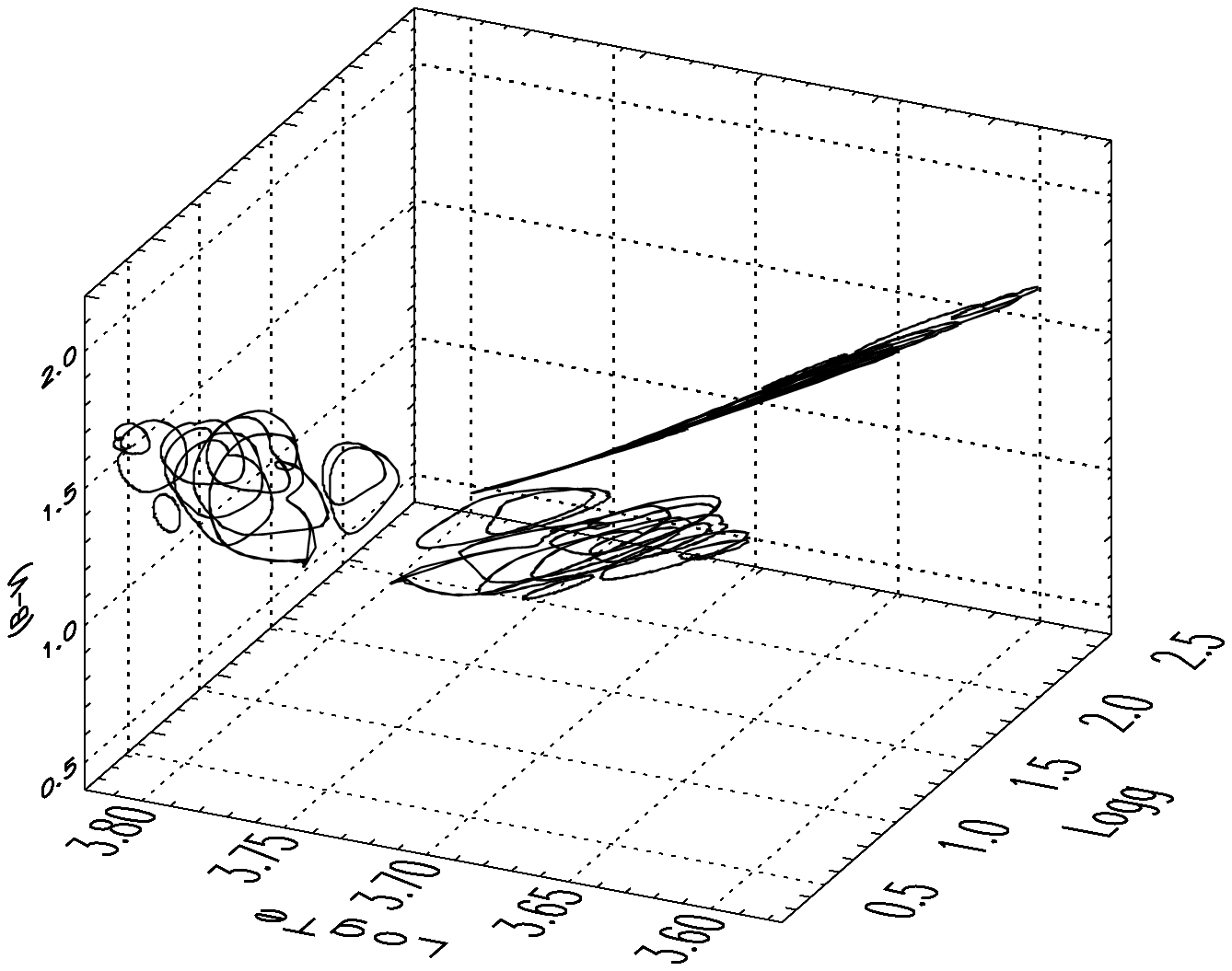,height=7cm}
\psfig{figure=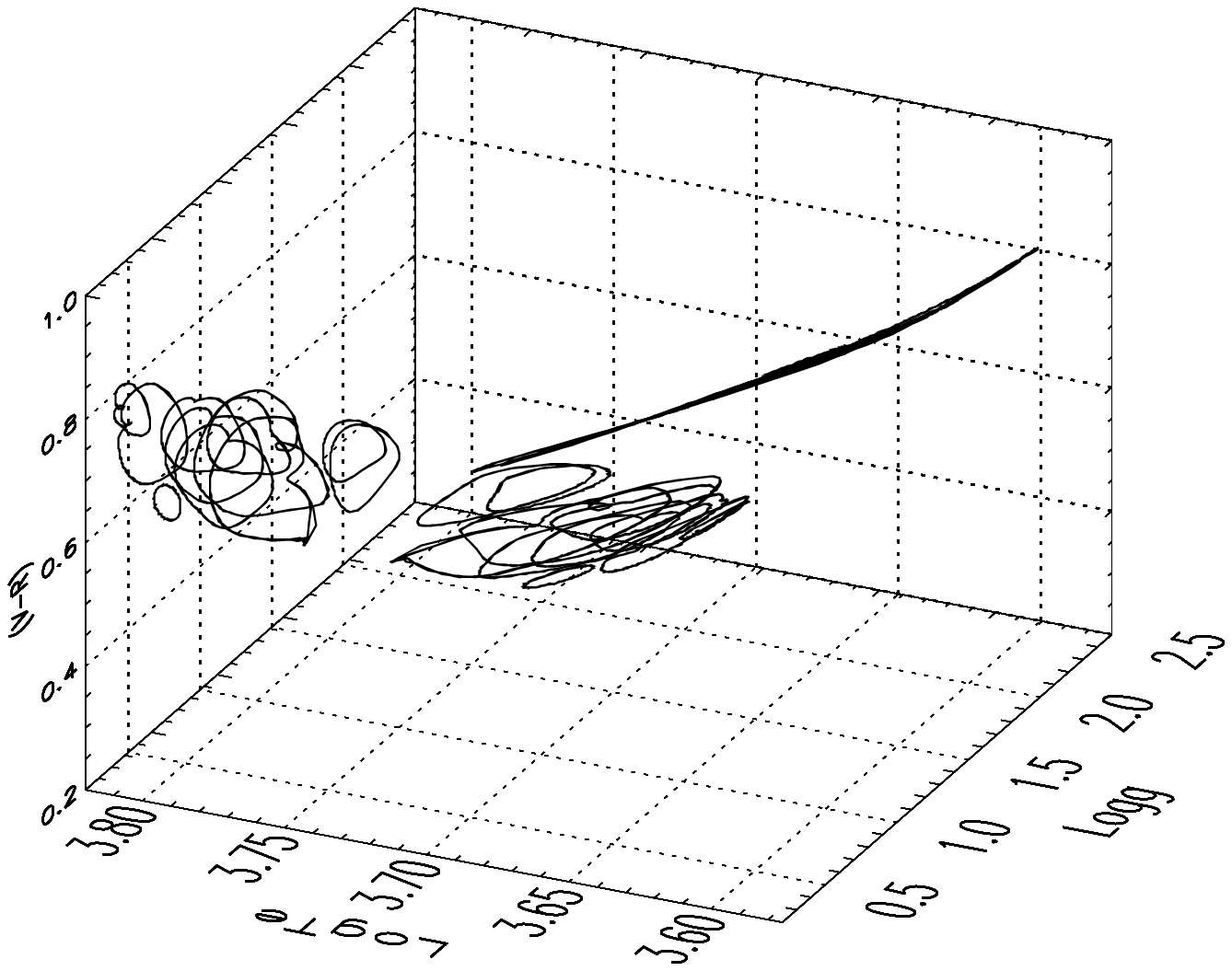,height=7cm}
}
\caption{Top plot: surface covered by the sample of theoretical models 
adopted in this investigation in the 3D space 
$[(B-V),log{T_e},log{g_{eff}}]$.  
In order to make clear the dependence of $(B-V)$ colors on both 
temperature and gravity the loop performed by each variable in the 
$(B-V)-log g_{eff}$ plane (left panel) and in the $(B-V)-log{T_e}$ 
plane (right panel) are also plotted. 
Bottom plot: same as above, but in the 3D space 
$[(V-R),log{T_e},log{g_{eff}}]$.
\label{fig5}}
\end{figure}

Even though previous results supply useful suggestions for improving 
the accuracy of radius measurements, the collection of both NIR and 
optical data for a large Cepheid sample is not a trivial observational 
effort. As a consequence, we decided to improve the approach suggested by 
RBMR for evaluating the $\Delta{B}$ term.  
Since $\Delta B$ is the area of the loop performed by each variable in 
the $S_V$-color plane, the idea is to compute $S_V$ along the pulsation 
cycle directly from observations. However, the surface brightness 
depends on both $T_e$ and $g_{eff}$, and therefore two relations 
should be inverted for deriving $S_V$: 

\begin{eqnarray}
C_1 &=& f~(T_e, g_{eff}) \nonumber \\
C_2 &=& g~(T_e, g_{eff}) \nonumber
\end{eqnarray}

\noindent
where $C_1$  and $C_2$ are two arbitrary colors. Unfortunately this 
problem does not admit a general solution over the whole parameter 
space, since the same color can be obtained for different pairs 
of $T_e$ and $g_{eff}$ values. This notwithstanding, it is 
still possible to find a local solution. 
Figure~\ref{fig5} show the surface covered by synthetic 
models in the 3D spaces $[(B-V),\log{T_e},\log{g_{eff}}]$ and 
$[(V-R),\log{T_e},\log{g_{eff}}]$ respectively.  
Data plotted in these figures show quite clearly that theoretical 
models populate a well-defined region of the quoted space. Therefore by 
performing a $4^{th}$ degree polynomial fit to the data it is possible 
to invert the two relations governing the $(B-V)$ and the $(V-R)$ colors as a 
function of temperature and gravity. The results of the polynomial 
approximations are presented in the appendix.

\begin{figure}
\psfig{figure=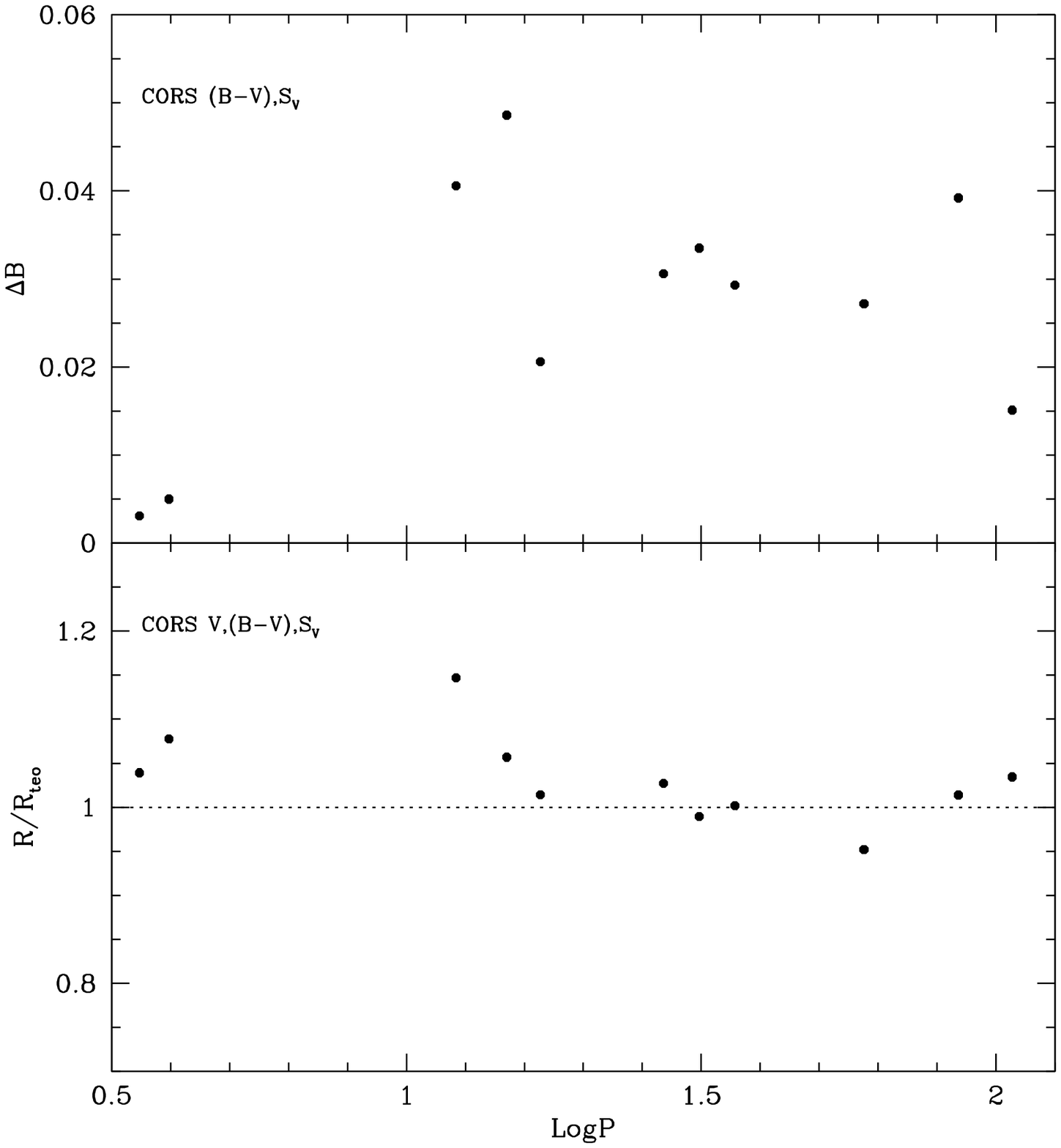,height=9cm}
\caption{Top panel: \db~ terms as a function of the logarithmic period 
obtained by adopting the revised CORS method and the theoretical 
calibration of the surface brightness. Bottom panel: similar to top 
panel but refers to the ratio between "computed" and "theoretical" 
radii. \label{fig6}}
\end{figure}

\begin{figure}
\psfig{figure=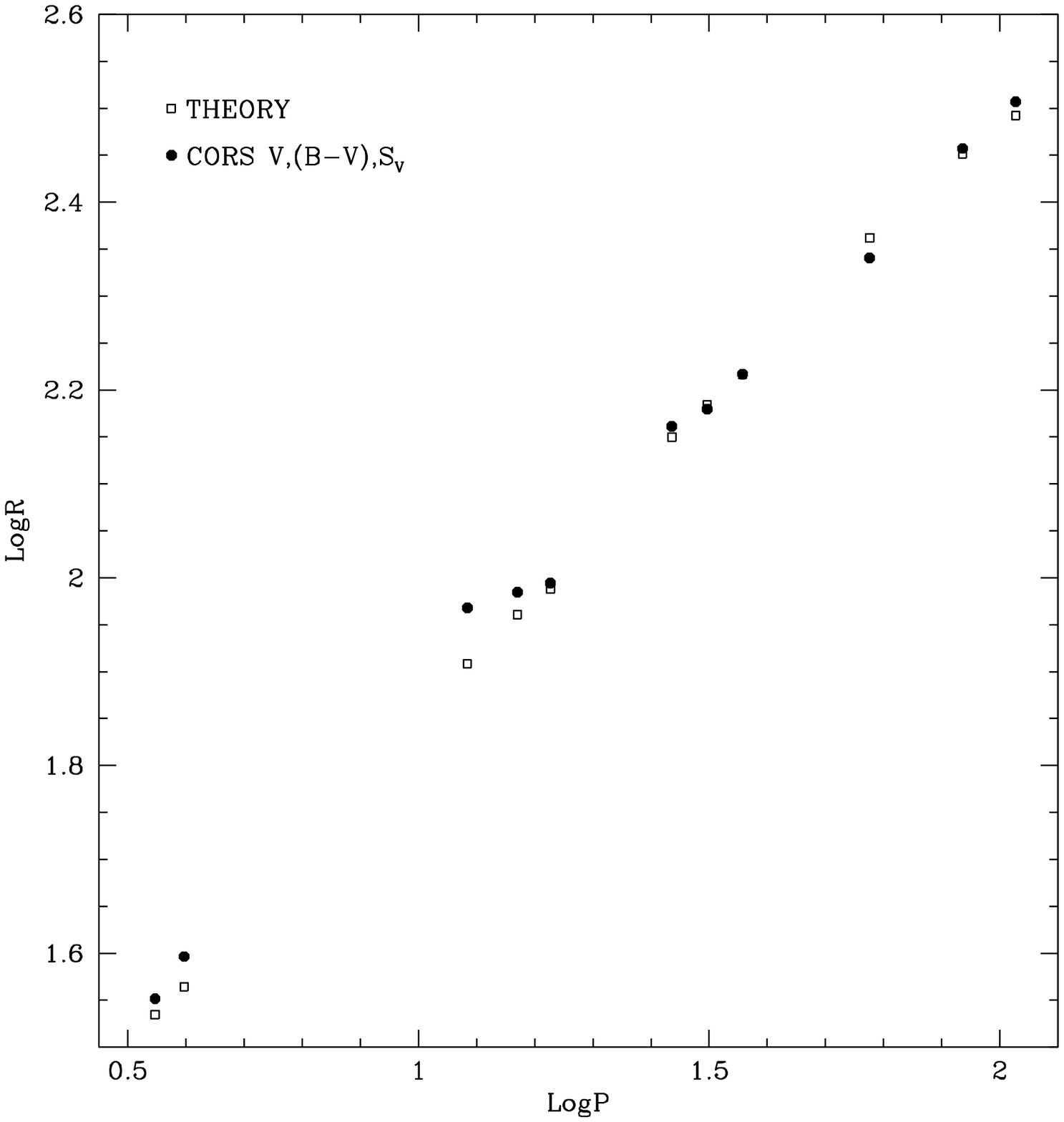,height=9cm}
\caption{Period-Radius relation in a log-log plane. The radius estimates
have been obtained by adopting the revised CORS method and the theoretical
calibration of the surface brightness. See text for further details. 
\label{fig7}}
\end{figure}

On the basis of these relations we can estimate the surface brightness 
$S_V$ directly from the following equation:

\begin{equation}
S_V=const.-10 \log {T_e} - BC 
\end{equation}

\noindent 
where the symbols have their usual meaning. The constant term 
depends on the photometric system used, but in our application 
it is not relevant, since the surface brightness in Eq.~\ref{eqdb} 
appears as a derivative. \\
By taking into account this new theoretical calibration we applied once
again the CORS method to the sample of synthetic models for estimating 
the Cepheid radii. Fig.~\ref{fig6} and \ref{fig7} show the results 
of these calculations.  
Data plotted in these figures support the evidence that:\\   
1) the theoretical calibration of the surface brightness we developed
is intrinsically correct. In fact, the discrepancy 
between "computed" and "theoretical" Cepheid radii is systematically 
smaller than 7\%. The only exception to this behavior is the radius 
of the model at 7\msun~ and $T_e=5300$ K which shows a well-defined  
bump along the rising branch. This evidence suggests that CORS 
estimates of Bump Cepheid radii could be affected by systematic 
errors. However, data plotted in Fig. 1 show that the 
outermost layers of this model undergo sudden gravity changes 
close to the bump phases. During these pulsation phases the assumption 
of hydrostatic equilibrium is no longer valid and therefore both the 
bolometric corrections and the colors obtained by adopting static 
atmosphere models should be regarded as suitable average estimates  
of the actual properties. It can be easily shown (Bono 1994) that this 
limit is mainly due to the effective gravity, since this quantity is 
estimated by assuming both radiative and hydrostatic equilibrium. 
The effective temperature only depends on the assumption of radiative 
equilibrium but the departures from the radiative equilibrium are,  
under the typical conditions of a pulsation cycle, quite small. 
These leading physical arguments suggest that the bump phases should 
be neglected in the CORS analysis.\\  
2) In comparison with "theoretical" radii the "computed" radii 
do not show any systematic shift. This result suggests that the 
\db~ terms evaluated by adopting the theoretical calibration 
-based on the polynomial approximations of $T_e$ and $g_{eff}$ 
in the color-color plane $[(B-V)-(V-R)]$ and of the BC in the 
[$\log T_e - \log g_{eff}$] plane- are more accurate than 
the \db~ terms obtained by means of the empirical calibration.\\  
3) Accurate radius determinations can be obtained by adopting photometric 
data in three optical bands. 

Finally, we mention that following a referee's suggestion we applied the 
revised CORS method to theoretical curves which mimic real observations. 
In particular, we performed several numerical experiments by sampling the 
theoretical $M_V$, $(B-V)$, $(V-K)$, and radial velocity curves with 
20-30 phase points randomly distributed along the pulsation cycle. 
To account for observational uncertainties the points were spread 
out by assuming typical photometric and spectroscopic errors and 
then fitted with up to 7 Fourier terms. 
Interestingly enough, we find that radius estimates based on these 
curves still present a discrepancy $\le$ 7\% when compared with theoretical 
radii. Therefore this uncertainty can be assumed as a plausible upper 
limit to the intrinsic accuracy of radius determinations based on the 
revised CORS method.

\section{Conclusions}

In this paper we present the results of a detailed investigation 
on the intrinsic accuracy of radius estimates obtained by adopting 
both the revised CORS method and the pure BW method. In order to 
avoid systematic errors the numerical experiments were performed 
by adopting theoretical observables -light, color, and radial 
velocity variations- predicted by nonlinear, convective 
models of classical Cepheids at solar chemical composition. 
The main findings of this analysis are the following:\\ 
1) the revised CORS method and the pure BW method applied to 
to NIR data provide radius estimates characterized on average 
by the same accuracy. However, the former method supplies more 
accurate radius determinations than the latter one when applied 
to optical bands.\\  
2) In agreement with current empirical evidence (Laney \& Stobie 1995) 
the PR relations obtained by adopting theoretical predictions are 
affected by the "photometric drift", i.e. the slope becomes steeper 
when moving from optical to NIR bands. Thus suggesting that at fixed 
period radius determinations based on NIR/optical bands 
overestimate/underestimate "true" radii. 
 
At the same time, in order to develop a method which can be applied 
to a large sample of Cepheids, we provided a new theoretical 
calibration of the $\Delta B$ term included in the revised CORS 
method. On the basis of this new calibration we find that the 
computed radii are affected by a discrepancy when compared with 
theoretical radii that is $\le$ 7\%. Moreover and even more 
importantly, we also find that computed radii based on optical 
bands do not show any systematic difference with theoretical 
radii.   
  
Obviously before any firm conclusion on the accuracy of the current
Cepheid PR relations can be reached, this method should be applied
directly to empirical data. However, the main interesting feature of
the current calibration is that it only relies on theoretical models. 
Therefore the comparison between theory and observations can allow 
us to supply tight constraints on the systematic uncertainties which 
affect radius estimates such as metallicity, reddening, and 
microturbulence velocity. 

Finally, we mention that direct measurements of Cepheid angular 
diameters through optical interferometry are becoming available
(Nordgreen et al. 2000, and references therein). In the near 
future, new and more accurate interferometric data can allow us 
to assess on a firm physical basis the calibration of the CORS method. 
At the same time, it is worth emphasizing that the development of a
homogeneous theoretical framework to be compared with new empirical
data can also supply sound suggestions on the plausibility and the
accuracy of the physical assumptions adopted for constructing both
pulsation and atmosphere models.

\begin{acknowledgements}
It is a pleasure to thank J. Lub as a referee for his detailed 
suggestions and helpful comments on an early draft of this paper.  
\end{acknowledgements}

\appendix 
\section{Polynomial approximations}

\begin{table*}[hbt]
\caption{Coefficients for the polynomial fits described in the Appendix. 
\label{coef}}
\begin{tabular}{rrrrrrrrr}
\hline
\hline\noalign{\smallskip}
$a_0$ & $a_1$ & $a_2$ & $a_3$ & $a_4$ & $a_5$ & $a_6$ & $a_7$ & $a_8$ \\
3.896 & -0.012 & 0.034 & -0.199 & 0.270 & -0.319 & -1.249 & 0.831 & 0.032 \\ 
\hline\noalign{\smallskip}
$b_0$ & $b_1$ & $b_2$ & $b_3$ & $b_4$ & $b_5$ & $b_6$ & $b_7$ & $b_8$ \\
0.963 & -168.126 & 124.432 & 251.995 & -5.932 & -147.224 & -80.090 & -14.874 & 75.675 \\ 
\hline\noalign{\smallskip}
$c_0$ & $c_1$ & $c_2$ & $c_3$ & $c_4$ & $c_5$ & $c_6$ & $c_7$ & $c_8$ \\
-81.064 &35.986&-3.827&-286.189&157.642&-21.706&104.568&-57.152&7.809\\
\noalign{\smallskip}
\hline
\hline
\end{tabular}
\end{table*}

The $4^{th}$ degree polynomial fit to effective temperature, effective 
gravity, and bolometric corrections mentioned in \S 4.3 are the following:  

\begin{eqnarray}
\log {T_e} &= &a_0+a_1(B-V)+a_2 {(B-V)}^2   \label {rel1} \\
           &+  & a_3 (V-R)+a_4 (B-V) (V-R)    \nonumber \\
           &+  & a_5 {(B-V)}^2 (V-R)+a_6 {(V-R)}^2 \nonumber \\
      	   &+  & a_7 (B-V) {(V-R)}^2+a_8 {(B-V)}^2 {(V-R)}^2 \nonumber
\end{eqnarray}

\begin{eqnarray}
\log {g_{eff}} &= &b_0+a_1(B-V)+b_2 {(B-V)}^2  \label {rel2} \\
           &+  & b_3 (V-R)+b_4 (B-V) (V-R)    \nonumber \\
           &+  & b_5 {(B-V)}^2 (V-R)+b_6 {(V-R)}^2 \nonumber \\
      	   &+  & b_7 (B-V) {(V-R)}^2+b_8 {(B-V)}^2 {(V-R)}^2 \nonumber
\end{eqnarray}

\begin{eqnarray}
BC &= & c_0 + c_1 \log T_e + c_2 \log T_e^2  \label {rel3} \\
           &+ &  c_3  \log g_{eff} + c_4  \log T_e \log g_{eff} \nonumber \\
           &+ &  c_5 \log T_e^2 \log g_{eff} + c_6  \log g_{eff}^2 \nonumber \\
   &+ &  c_7 \log T_e \log g_{eff}^2 + c_8  \log{T_e}^2 \log g_{eff}^2 \nonumber
\end{eqnarray}

\noindent
the coefficients $a_i, b_i, c_i$  of the previous relations are listed in 
Table~\ref{coef} and the other symbols have their usual meaning.
Note that the r.m.s. of the previous relations are 0.0008, 0.04 and 0.003 
respectively.

{}

\end{document}